\documentstyle[aps,prd]{revtex}
\newcommand{\tend}{\rightarrow}

\newcommand{\bea}{\begin{eqnarray}}
\newcommand{\eea}{\end{eqnarray}}
\newcommand{\bfig}{\begin{figure}}
\newcommand{\efig}{\end{figure}}
\newcommand{\barr}{\begin{array}}
\newcommand{\earr}{\end{array}}
\newcommand{\bean}{\begin{eqnarray*}}
\newcommand{\eean}{\end{eqnarray*}}
\newcommand{\bdm}{\begin{displaymath}}
\newcommand{\edm}{\end{displaymath}}
\newcommand{\beq}{\begin{equation}}
\newcommand{\eeq}{\end{equation}}

\newcommand{\meio}{\frac{1}{2}}

\newcommand{\del}{\partial}

\newcommand{\eps}{\epsilon}

\newcommand{\espc}{\hspace{1cm}}
\newcommand{\cP}{{\cal P}}

\begin{document}

\title{Chaos and Fractals in Geodesic Motions Around a Non-Rotating
Black-Hole with an External Halo}
\author{Alessandro P. S. de Moura\footnote{email:
sandro@ifi.unicamp.br}}
\address{Instituto de F\'{\i}sica Gleb Wataghin, UNICAMP, Brazil}
\author{Patricio S. Letelier\footnote{email:
letelier@ime.unicamp.br}}
\address{Instituto de Matem\'atica, Estat\'{\i}stica e Ci\^encia da
Computa\c{c}\~ao, Departamento de Matem\'atica Aplicada, UNICAMP,
Brazil}
\maketitle

\begin{abstract}
We investigate the occurrence chaos in the escape of test particles moving in the
field of a Schwarzschild black hole surrounded by an external
halo. The motion of both material particles and zero rest mass
particles is considered. The chaos is characterized by the fractal
dimension of boundary between the basins of the different escapes,
which is a topologically invariant characterization. We find chaos
in the motion of both material particles and null geodesics.
\end{abstract}

\section{Introduction}

The study of chaos in dynamical systems with unbounded orbits is
relatively recent\ \cite{ott}. The characterization of chaos for this
class of problems is different from that used for bounded dynamics,
which is based upon the destruction of KAM tori. One
of the most important situations with unbounded motion is that
of the escape of particles from a certain region; this problem is
closely related to scattering, the difference between the two being
essentially the choice of the initial conditions. Escapes have been
studied for several systems: galactic dynamics\ \cite{contopoulos},
nonlinear oscillations\ \cite{thompson,stewart}, two-dimensional
conservative mappings\ \cite{christiansen} and inflationary cosmology\
\cite{cornish} are only a few examples.

	In this paper, we study the dynamics of unbounded
orbits of test particles (including zero rest mass particles) in a
general-relativistic vacuum static axisymmetric system consisting of a
non-rotating black hole surrounded by an external multipolar halo; in
particular, we are interested in the escape properties of these
systems. If a system has two or more physically well-defined escapes
for a given set of parameters of the metric (for instance, regions
where a particle runs away to infinity, or where it falls into
an event horizon), then the escape it chooses is a function of its
initial conditions; when this function has a fractal structure, we
have a well-defined kind of chaos, and the correspondent fractal
dimension gives a good quantitative characterization of the chaos,
besides having a simple physical interpretation as a measure of the
sensitivity to initial conditions (see section III). Since the fractal
nature of the boundary between the escapes as well as its associated
fractal dimension are topological invariants, they are independent of
the choice of the space-time coordinates; this assures the
meaningfulness of this characterization for General Relativity.

	This paper is organized as follows: in section II we review
the Weyl form of the general vacuum static axisymmetric metric and
some of its properties; in section III we define the box-counting
dimension and discuss its physical significance; in section IV, we
investigate the basins of escape in the motion of material particles
for some choices of static axisymmetric metrics, and show numerically
the existence of chaos; in section V, we show that null geodesics are
regular (non-chaotic) in the field of a dipolar halo (plus the black
hole), but chaos arises if we add multipole moments of higher order to
the halo; and in section VI we summarize our results and draw some
conclusions.

\section{The Weyl Metric}

	Throughout this article, we use the Weyl metric to describe a
general static axisymmetric space-time\cite{robertson}:
\beq
ds^2 = e^{2\psi}dt^2 - e^{-2\psi}\left[e^{2\gamma}(dr^2 + dz^2) +
	r^2d\phi^2\right] ,
\label{weyl}
\eeq 
where $r$ and $z$ are the radial and axial coordinates, and
$\phi$ is the angle about the $z$ axis, which is the axial symmetry
axis. Throughout this article, we will use units such that $c=1$ and
$m=1$, where $m$ is the mass of a test particle moving in spacetime
(\ref{weyl}). $\psi$ and $\gamma$ are functions
of $r$ and $z$ only. In these coordinates, the vacuum Einstein
equations reduce to: 
\beq \triangle\psi \equiv \frac{\del^2\psi}{\del r^2} + 
\frac{1}{r}\frac{\del\psi}{\del r} + \frac{\del^2\psi}{\del z^2}
= 0 ;
\label{eqnu}
\eeq
\beq
d\gamma = r\left[\left(\frac{\del\psi}{\del r}\right)^2 -
	\left(\frac{\del\psi}{\del z}\right)^2\right]dr +
	2r\frac{\del\psi}{\del r}\frac{\del\psi}{\del z}dz .
\label{eqgamma}
\eeq

The first expression is just Laplace's equation in cylindrical
coordinates; the second equation is a quadrature whose integrability
is automatically guaranteed by eq.(\ref{eqnu}).

The metric (\ref{weyl}) is independent of the time $t$ and of the
symmetry angle $\phi$. From this we obtain the two constants of motion
$E$ (energy) and $L_z$ (projection of the angular momentum on the
symmetry axis), that are conserved along the trajectories of test
particles in the metric (\ref{weyl}). They are given by: 
\label{cons}
\beq 
E \equiv p_t = g_{tt}\dot{t}; 
\eeq 
\beq 
L_z \equiv p_\phi = g_{\phi\phi}\dot{\phi} ,
\eeq 
where the dot denotes differentiation with respect to an affine
parameter. The only independent dynamical variables are thus $r$,
$z$ and their momenta $p_r$ and $p_z$: the time evolution of $t$ and
$\phi$ are given by the quadratures 
above. This means that the dynamical system corresponding to the
motion of test particles in the Weyl metric has only two degrees of
freedom.

To proceed further, it is convenient to define the prolate spheroidal
coordinates $u$ and $v$ by:
\beq
z = uv ;
\eeq
\beq
r^2 = (u^2-1)(1-v^2) ;
\eeq
\bdm
u\geq 1; \espc -1\le v\le 1 .
\edm

In these coordinates, equation (\ref{eqnu}) is separated, and a
general solution is obtained in a series of products of Legendre
polynomials and zonal harmonics\ \cite{eqdif}. The solution we are
interested in is: 
\beq
\psi = \meio\ln\left(\frac{u-1}{u+1}\right) +
	\sum_{n=1}^\infty a_nP_n(u)P_n(v) .
\label{mexp}
\eeq
The first term represents a Schwarzschild black hole, and the terms
under the summation sign are multipolar contributions from an external
halo.

Using the coordinates $u$ and $v$ and the expression (\ref{mexp}) for
$\psi$, $\gamma$ can be obtained from a straightforward integration of
eq. (\ref{eqgamma}); the constant of integration is chosen so as to
avoid conical singularities on the $z$ axis, by imposing $\gamma=0$
for $r=0$ and $|z|>1$.

In this article, we are interested in the multipole contributions only
up to the octopole term ($n=3$ in eq.(\ref{mexp})). Redefining the
coefficients in the expansion, we can write $\psi$ as:
\bea
\psi & = & \meio\ln\left(\frac{u-1}{u+1}\right) -
	  Duv + (Q/6)(3u^2-1)(3v^2-1) + \nonumber \\
    &   & (O/10)uv(5u^2-3)(5v^2-3) ,
\label{expan}
\eea
 where $D$, $Q$ and $O$ are defined as the dipole, quadrupole and
octopole moments. Now an explicit expression for $\gamma$ may be found
by direct integration. Since the expressions are quite cumbersome and
not particularly illuminating, we will not write them here; they can
be found in \cite{let1,let2}. We only observe that due to the
nonlinearity of Einstein's equation, reflected in this case in
eq. (\ref{eqgamma}), there are nonlinear terms of interaction between
the multipole terms in $\gamma$: the gravitational field due to the
different terms in the expansion (\ref{mexp}) is not simply the
superposition of the fields due to each term separately.

We finish this discussion by observing that the coordinates $u$ and
$v$ describe the metric only outside the black hole; in these
coordinates, the event horizon is given by the segment $r=0$, $|z|\leq
1$. Since we are interested only in the motion of particles outside
the event horizon, this singular behavior of the coordinate
transformation $(r,z) \tend (u,v)$ will not concern us here.

\section{Fractal Basin Boundaries}

We now review briefly some basic concepts on fractals in dynamical
systems with escapes; a complete discussion is found in \cite{ott}.

In order to simplify the discussion, we will suppose we have a region
in the phase space of the system (denoted the {\em inner region})
which has two distinct escapes, denoted by 1 and 2 (the generalization
for a higher number of escapes is straightforward). By {\em escape} we
mean a route that allows the particle to leave the inner region
permanently. The particular escape chosen by a particle is dependent
on the initial conditions of that particle. For a given dynamical
system and a given energy, the set of points in phase space which
correspond to initial conditions such that the particle chooses escape
1 is the {\em basin} corresponding to escape 1; the basin
corresponding to escape 2 is defined analogously. A point in phase
space is defined to be a boundary point if every neighborhood of such
a point contains points belonging to both basins. The basin boundary
is the set formed by all the boundary points.

This system is chaotic if its basin boundary is fractal. Near a
fractal basin boundary the points belonging to the different basins
are mixed in a very complex way, down to arbitrarily small scales. If
we draw a plot of the basins with a finite resolution, and amplify
a region containing a fractal boundary, then no matter how much
we amplify it, we will always find complex structures of intermixing
points of both basins. This implies a strong dependency on the initial
conditions near a fractal basin boundary.

If a system has a fractal basin boundary, then it has a fractal set of
unstable ``eternal'' orbits, that never escape in the past and in the
future (orbits that have never entered nor will ever leave the inner
region), called the {\em invariant set}. The basin boundary is formed
by trajectories belonging to the stable manifold of the invariant set,
that is, by trajectories that never escape in the future (``trapped''
trajectories); the unstable manifold is formed by orbits that do not
escape for $t\tend -\infty$. The invariant set, as well as its stable
and unstable manifolds, are sets of zero measure within the phase
space. Notice that in general not all the eternal trajectories belong
to the invariant set: if the system has a stable periodic orbit for
energies above the escape energy, then orbits near this one will also
be eternal, and they form a non-zero-measure set of eternal orbits
that are not part of the invariant set. These stable orbits have
islands of regular behavior surrounding them, where the phase-space
flow is integrable and confined to invariant tori.

We note that, for a system (at a given energy) to have a fractal basin
boundary, it needs not only to be non-integrable, but also it must be
such as to allow the presence of such fractal set of trapped
trajectories. In other words, the potential must be such that the
particle can bounce back and forth many times before it escapes, if
the system is to have a fractal basin boundary.

The presence of a fractal invariant set is the result of transversal
crossings of the stable and unstable manifolds of isolated unstable
periodic orbits that lie near the openings of the potential, the
so-called Lyapunov orbits\ \cite{contopoulos}. The basin boundaries
between the different escapes are the stable manifolds of the Lyapunov
orbits. The homoclinic and heteroclinic crossings imply a horseshoe
symbolic dynamics, which is responsible for the chaos and the fractal
character of the basin boundaries. The horseshoe dynamics results also
in the existence of a dense set of countable unstable periodic orbits,
that must thus exist if the system has a fractal basin boundary.

To give a quantitative measure of the sensibility to initial
conditions of a system with a fractal basin boundary, we define the
box-counting dimension of the boundary as follows: let two points
chosen randomically in a region of the phase space be separated by a small
distance $\eps$; it can be shown\cite{ott} that the probability that
the two points belong to different basins scales as: 
\beq 
P(\eps) \propto \eps^{D-d} ,
\label{dim}
\eeq 
where $D$ is the (integer) dimension of the region where the
ensemble of points was chosen; and $d$ is the (possibly non-integer)
dimension of the intersection of the basin boundary with this
region. If the boundary is non-fractal, then $d=D-1$, while if the
boundary is fractal, we have $d>D-1$. By choosing randomly a large
number of points in a region of the phase space for a certain fixed
$\eps$, we can calculate $P(\eps)$ numerically, and doing this for
several values of $\eps$, we can calculate the fractal dimension $d$;
this is the method we use in this article.

If our system has a stable periodic orbit for energies above escape,
then as we said above it has a set of positive measure of non-escaping
regular orbits confined to tori in phase-space. Escaping orbits that
come close to this set stay in its neighborhood for a long time before
leaving; in other words this set is ``sticky'' \cite{mapa}. This
complicates the task of calculating the box-counting dimension, for it
demands a greater integration time. We also observe that for this same
reason, the boundary between the escaping and the non-escaping regular
orbits does not have a well-defined box-counting dimension.

We observe that since both the fractal structure of the basin boundary
and its dimension $d$ are topological invariants of the dynamical
system, they are valid characterizations of chaos in general
relativity.

\section{Dynamics of Material Particles}

In this section, we investigate the movement of material test
particles in the metric (\ref{expan}) for some choices of the
multipole moments $D$, $Q$ and $O$. Important properties of the
dynamics can be understood by means of the ``effective potential''
associated with this metric\ \cite{gravitation}.

Besides the energy and the $z$ component of the angular momentum,
there is another quantity which is conserved along the trajectory of a
test particle, namely its rest mass. Since we are using units such that the
rest mass is unity, the conserved quantity is:
\beq
g^{\mu\psi}p_\mu p_\psi = E^2g^{tt} + L_z^2g^{\phi\phi} + 
	f(\dot{r}^2 + \dot{z}^2) = 1 ,
\label{consm}
\eeq
where $f=-g_{rr}=-g_{zz}=e^{2(\psi-\gamma)}$. The boundary of the region
in the configuration space which is accessible to the particle is found by
setting $\dot{r}=\dot{z}=0$:
\beq
E^2g^{tt} + L_z^2g^{\phi\phi} - 1 = 0.
\eeq

The ``effective potential'' $V(r,z)$ is then given by:
\beq
V(r,z) \equiv E^2 = \frac{1 - L_z^2g^{\phi\phi}}{g^{tt}} ;
\eeq
Substituting for the Weyl metric (\ref{weyl}), we have:
\beq
V(r,z) = e^{2\psi}\left(1 + \frac{e^{2\psi}L_z^2}{r^2}\right) ;
\label{potm}
\eeq
The region on the $rz$ plane accessible to the particle is given by
$V(r,z)\le E^2$.
Using equation (\ref{expan}) for $\psi$, we thus have an
expression for the effective potential in terms of the multipole
moments. We note that $V$ depends only on $\psi$, and not on $\gamma$.

The equations of motion for the test particles is:
\beq
\ddot{x}^\mu + \Gamma_{\alpha\beta}^\mu \dot{x}^\alpha \dot{x}^\beta =
0 ,
\eeq
and using eqs. (\ref{cons}) and (\ref{weyl}) we cast them in the
convenient form:
\label{eqmot}
\beq
\ddot{r} = -\frac{1}{2f}\left[g^{tt}_{,r}E^2 + g^{\phi\phi}_{,r}L_z^2
           + f_{,r}\left(\dot{r}^2-\dot{z}^2\right)
	   + 2f_{,z}\dot{r}\dot{z}\right] ;
\eeq
\beq
\ddot{z} = -\frac{1}{2f}\left[g^{tt}_{,z}E^2 + g^{\phi\phi}_{,z}L_z^2
           + f_{,z}\left(\dot{z}^2-\dot{r}^2\right)
	   + 2f_{,r}\dot{r}\dot{z}\right] .
\eeq

The evolution of $t$ and $\phi$ is given by the quadratures
(\ref{cons}).

In the following subsections, we will analyze the dynamics for some
interesting choices of $D$, $Q$ and $O$. For bounded trajectories,
this system was shown through Poincar\'e sections to be chaotic\
\cite{let1,let2}.

\subsection{Dipole Potential}

If we make $Q=O=0$ in (\ref{expan}), we have a pure dipole field
together with a Schwarzschild black hole. Bounded trajectories of this
system have been studied, and chaos has been found using Poincar\'e
sections\cite{let2}; the classical equivalent of this system can be
shown to be integrable, so the chaos is due to general relativistic
contributions to the dynamics. We shall now study this system in the
open regime, that is, with energies large enough as to allow them to
escape either to infinity or to the event horizon.

In figure 1 we show some contour levels of the effective potential
$V(r,z)$ for $D=3\times 10^{-4}$ and $Q=O=0$, with $L_z=3.0$. The first
feature we notice is the ``tunnel'' formed by the equipotential curves
for small values of $r$, that leads to the event horizon (remember
that in these coordinates, the event horizon is given by $r=0$ and
$|z|\le 1$). This is the route followed by particles that fall into
the black hole. We observe that $V$ is invariant under the
transformation $z\tend -z; \hspace{0.5cm} D\tend -D$.

If a particle has high enough energy ($E^2$ higher than about 0.94 for the
parameters of fig. 1), it can also escape to infinity. This is shown
clearly in fig. 1 by the opening that appears in the equipotentials
for high energies.

Now let us pick one specific value for the energy, for instance
$E^2=0.95$. At this energy, a particle's orbit can have three
outcomes: (1) escape into the black hole; (2) escape to infinity; and
(3) keep bouncing back and forth forever, and never leave the
inner region. The concave shape of the equipotential for this
energy suggests that we have stable trapped orbits and thus regions of
regular behavior in the phase space. This is indeed the case, as we
will see shortly.

To investigate the nature of the basin boundaries, we need a portrait
of the basins; to do this, we 
define a 2-dimensional section of the 3-dimensional energy shell of
the phase space that is accessible to the particle. For
$E^2=0.95$ we define this section as the set of initial conditions
with spatial coordinates lying on the segment given by $z=0$
and $15\le r\le 25$, with velocities given by:
\beq
\dot{r} = v\cos(\theta); \espc
\dot{z} = v\sin(\theta) ,
\label{vel}
\eeq
where 
\beq
v = (\dot{r}^2+\dot{z}^2)^{1/2} =
      \frac{1}{f}\left(1 - E^2g^{tt} - L_z^2g^{\phi\phi}\right) 
\eeq
is fixed by the conservation
equation (\ref{consm}), and $0\le\theta\le 2\pi$. This section is thus
a topological segment of a cylinder embedded within the phase space,
and we will denote it by $S$.  

To obtain numerically the intersection of the basins with this
section, we divide the intervals $15\le r\le 25$ and $0\le\theta\le
2\pi$ into 400 equal parts each; this defines a grid on $S$ composed
of $400\times 400$ points. For each of these points, we integrate
numerically the equations of motion for the dipole metric, and record
the outcome: if the trajectory falls into the black hole (numerically,
if $r$ becomes too small, or less than $0.5$ in this case), that
initial condition belongs to basin 1; if the trajectory escapes to
infinity (numerically, if $r$ or $z$ becomes too large, larger than
$60$ in this case), it belongs to basin 2; and if after a certain
proper time $\tau_{max}$ (in this case $\tau_{max}=10000$) the
trajectory chooses none of the two escapes above, then we admit that
it belongs to the set of ``trapped'' trajectories that never leaves
the confining region. We choose $\tau_{max}$ such that the set of
``trapped'' trajectories is well resolved for the scale of the grid we
use.

The results of this calculation are shown in fig. 2a. A black dot
means that the corresponding point $(r,\theta)$ in $S$ belongs to
basin 1; a white dot indicates that it belongs to basin 2; and a grey
dot means it belongs to the set of ``trapped'' trajectories. We notice
a complex Cantor-like mixing of basins, indicating that the structure
continues down to smaller scales. This is confirmed by the
amplification of a detail of figure 2a, shown in fig. 2b.  The area
covered by fig. 2b is about 10 orders of magnitude smaller than that
of fig. 2a, giving strong evidence that the basin boundary is indeed
fractal. We note that the set of trapped trajectories has as we
suspected a non-zero measure; this is clear from fig. 2a. As we
discussed in section III, we expect them to form a regular island in
phase space. The intersection of these orbits with a Poincar\'e
section should define closed curves. This can in fact be observed in
fig. 3a, which shows the intersection of some of these trapped orbits
with the surface of section $z=0$. The parameters are the same as in
fig. 2. The regular island shown in fig. 3a appears to be the only one
present in the system for these parameters.

To have a more precise and quantitative characterization of the
fractal structure seen in fig. 2, we proceed to the calculation of the
fractal dimension, as discussed in section III. The random points are
chosen in $S$, and for each point $(r,\theta)$ we find through numerical
integration to which of the basins it belongs, and then do the same
for two nearby phase space points given by $r+\eps$ and $r-\eps$ and
the same $\theta$. If all three points do not belong to the same
basin, then the point $(r,\theta)$ is considered an ``uncertain''
point, meaning that it lies close to a basin boundary. For a large
number $N$ of points randomly chosen in $S$, the fraction of uncertain
points is $f(\eps) = N'/N$, where $N'$ is the number of
uncertain points found in the sample of $N$ points. For
$N$ large enough, $f$ is proportional to $P$ in eq. (\ref{dim});
finding in this way $f$ for several values of $\eps$, a log-log
plot of $f(\eps)$  should give a straight line, and 
the basin boundary dimension $d$ is found by the angular coefficient
through eq. (\ref{dim}). For reasons explained in sec. III, we
calculate $d$ choosing points in a region that does not intersect the
trapped regular region. By getting rid in this
way of the ``stickiness'' of the regular region, we are able to obtain
a meaningful result for $d$.

The results are shown in fig. 3b. We have chosen $N$ such that
$N'>100$; this means a statistical uncertainty of 10\% in $f$. We see
that the points lie on a clearly defined straight line; the angular
coefficient is $\alpha=0.47 \pm 0.02$, which gives a
dimension of $d=1.53 \pm 0.02$, showing unambiguously that the
boundary is fractal. We remember that $d$ is the dimension of the
intersection of the basin boundary with the 2-dimensional section $S$;
the dimension of the basin boundary in the accessible 3-dimensional
space is $d+1$, and its dimension in the full 4-dimensional
phase-space is $d+2$.  We have calculated $d$ for some subregions of
$S$, and we have obtained always the same value to within the
statistical uncertainty, showing that the method is self-consistent
and the result is meaningful.

We have studied how the box-counting dimension varies as we change the
various parameters of the metric. If $D=0$, a particle needs an energy
$E$ higher than 1 to be able to escape to infinity. If $D\ne 0$, the
escape energy becomes less than 1, and depends on the angular momentum
$L_z$. We denote the escape energy by $E_0=E_0(L_z)$. The basin
boundary dimension $d$ is defined only for $E>E_0$. We have found that
for $E>1$, $d=1$ (to within the statistical error), and the motion is
regular. We have verified this result for several values of $L_z$ end
$E$, and three different values of $D$; this feature is also suggested
by the forms of the equipotentials for $E>1$. 

We have also investigated how $d$ varies with the dipole strength
$D$. In the limit $|D|\tend\infty$, we have a field dominated by the
dipole component; the geodesics defined by a pure dipole field are
integrable, and thus we expect $d$ to approach 1 for high values of
$D$. If we decrease $D$ enough, we end up reaching a value $D_0$ below
which the particle can no longer escape to infinity, and $d$ is no
longer well defined. Near $D=D_0$, with $D>D_0$, the opening of the
equipotential to the escape to infinity is small, and the particle is
likely to bounce more times before it escapes through this route than
in the case of higher values of $D$, and we accordingly expect the
chaos to be ``larger'' in this case, that is, $d$ to be larger. These
features are indeed verified in a plotting of $d$ versus $D$ for
$E^2=0.95$ and $L_z=3.0$, shown in fig. 4. For these values of $D$ and
$L_z$, we have $D_0\equiv 2.5\times 10^{-4}$. The system is 
regular ($d=1$) for $D\ge 9\times 10^{-4}$, $d$ reaches its highest
value of about 1.6 at $D=D_0$.

\subsection{Quadrupole Potential}

We next turn to the case $D=O=0$. This quadrupole field has a
reflection symmetry with respect to the $z$ axis: the metric is
unchanged by $z\tend -z$. For $Q>0$, the open equipotentials are
similar to the pure dipole case discussed above (except for the
aforementioned symmetry). A more interesting
choice is $Q<0$; in this case, as shown in fig. 5, there are two
different escapes to infinity, besides the escape into the event
horizon. In this case, we can choose an energy level such that the set
of trapped trajectories has zero measure, and therefore there are no
regular regions in phase space, as in the dipole case. As stated in
section III, this is desirable numerically, since the presence of
trapped trajectories increases very much the integration times needed. The
energy we have chosen is $E^2=0.97$, with $L_z=2.6$.

We proceed as we did for the dipole case. We choose initial conditions
in the segment $r=25.0$, $|z|<25.0$; the velocities are given by
(\ref{vel}). The results are in fig. 6a, with black dots denoting
trajectories that escape upwards, white dots denoting trajectories
that escape backwards, and grey dots denoting trajectories that fall
into the event horizon. Figure 6b shows an amplification of a very
small area of fig. 6a, and the absence of smoothness in the basin
boundary shows clearly its fractal character. The fractal dimension
was computed as described above, and the value we obtained was $d =
1.60 \pm 0.03$. We have calculated $d$ for other values of the energy
and angular momentum, and we found that, as opposed to the dipolar
halo system studied in the previous section, this system is chaotic
for $E>1$. In fact, we found that the
boundary is fractal for arbitrarily large values of the energy (for
$Q<0$), as far as we have been able to investigate; this is an
important difference between the dipolar and quadrupolar halos.

Since the basin boundary between the escapes is the stable manifold of
an invariant set, we have associated with the chaos in the choice of
the escape route a chaos in the escape time as well, as is well-known
in chaotic scattering. This happens because orbits starting from very
close initial conditions may make a different number of bounces before
escaping, leading to very different escape (proper) times. We have
illustrated this by finding numerically the escape proper times
$\tau_e$ for orbits starting from a fixed position $r=25$, $z=0$, for
several velocity angles $\theta$, as defined by eq. (\ref{vel}). We
plot $\tau_e(\theta)$ in fig. 7a. The ``spiked'' character of the
graph is striking, suggesting a fractal structure. This is confirmed
by fig. 7b, which shows that the function $\tau_e(\theta)$ has a
fractal set of singular points, where $\tau_e(\theta)$ goes to
infinity; this set is the intersection of the line of initial
conditions with the basin boundary. We observe that the escape time
$\tau_e$ is to some extent arbitrary, because it depends on where we
stop the integrations of the trajectories before we consider them to
have escaped. However, the fractal structure seen in fig. 7 is
topological, and is not affected by this choice.

In order to gain more insight into the fractal structure of the basin
boundary and its related complex dynamics, we now define a surface of
section in phase space denoted by $\cP$ and given by $\dot{z}=0$, for
a given $E$ and $L_z$. We define $I_n$ ($n\ge 1$) as the set of points
on the surface of initial conditions $S$ that generate orbits that
cross $\cP$ at least $n$ times in the negative direction (that is,
satisfying $\ddot{z}<0$) before escaping. Obviously $I_n$ is a subset
of $I_k$ if $k<n$, and we have that:
\beq
I_1 \supset I_2 \supset I_3 \supset \cdots .
\eeq

For systems without regular trapped orbits (such is the case of the
quadrupole field with the parameters of fig. 7), the basin boundary is
given by $\lim_{n\tend +\infty}I_n$. If we let $I_n$, with $n$ being a
negative integer, denote the set of points in $S$ corresponding to
past-directed orbits which cross $\cP$ in the negative direction at
least $|n|$ times, then the unstable manifold of the repellor set is
analogously given by $\lim_{n\tend -\infty}I_n$. Defining now the set
$R_n = I_n \cap I_{-n}$, the repellor set is given by $\lim_{n\tend
+\infty}R_n$. This simply states that the repellor set is the
intersection of its stable and unstable manifolds.

The mechanism of the construction of the fractal basin boundary by the
dynamics of the system may be followed by examining the sets $I_n$
($n>0$). Figs. 8a, 8b and 8c show $I_1$, $I_2$ and $I_3$ respectively,
using the same grid as that of fig. 7a. As $n$ increases, the
structure of $I_n$ becomes more and more complicated: every step
$I_n\tend I_{n+1}$ results in taking from $I_n$ increasing numbers of ever
thinner strips, intercalatedly. We recognize this as the mechanism for
the construction of a Cantor set, in the limit $n\tend +\infty$. 

We can follow in the same way the construction of the invariant set
itself: in fig. 8d we show $R_2$; compare this with fig. 8b.

The Newtonian system equivalent to the multipole field we are dealing
with is given by the Hamiltonian:
\beq
H = \meio\left(p_r^2 + p_z^2\right) + V(r,z),
\label{hamclass}
\eeq
where $V(r,z)$ is the effective potential:
\beq
V(r,z) = \frac{L_z^2}{2r^2} - \frac{1}{r} + \psi(r,z),
\label{effpotclass}
\eeq
and $\psi$ is given by (\ref{expan}). For the dipole field ($Q=O=0$),
as we mentioned before, the Hamiltonian (\ref{hamclass}) is
integrable. For the quadrupole field, however, it is not\cite{let1},
so we expect to have a fractal basin boundary for this case as
well. Figure 9 shows some level contours of $V$ with $L_z=2.6$ and
$Q=-4\times 10^{-6}$, $D=O=0$; these are the same parameters we have
used for the relativistic case. For this negative value of $Q$, we have two
escapes (for $Q>0$, there is only one escape). Since we want to
compare the Newtonian and the relativistic cases, we choose the energy
to be $E-1$, where $E$ is the energy we used in the relativistic case,
which gives -0.0151. We proceed as in the relativistic case to calculate
the basin boundary dimension. The initial conditions are chosen in
the segment $|z|<5$, $r=13$. The result is
$d = 1.64 \pm 0.02$, which is roughly the
same (actually a little larger) value we obtained for the relativistic case.

\subsection{Quadrupole + Octopole Potential}

The last case we investigate in this section is the field formed by
the superposition of the quadrupole and octopole components, $D=0$
with $Q,O \neq 0$. The octopole term breaks the reflection symmetry of
the quadrupole potential, as can be seen in fig. 10, which shows some
level contours of the effective potential for $D=0$, $Q=-4\times
10^{-6}$, $O=-1\times 10^{-7}$ and $L_z=2.6$. We still have three
escapes as in the previous case, but the equipotentials are distorted,
and are no longer symmetrical with respect to the $z=0$ axis. We
select the energy $E^2=0.97$, and pick the initial conditions on the
segment $|z|<20$, $r=20$. Using these parameters, we have calculated
the basin boundary dimension, and found $d=1.59\pm 0.02$, which is
practically the same value obtained for the pure quadrupole field
$O=0$. If we calculate $d$ for the pure octopole field $O=-10^{-7}$,
$Q=0$, keeping the other parameters fixed, we find $d=1.70\pm 0.02$,
which is {\em larger} than the value obtained for the mixed field.
This is a somewhat surprising result, and it shows that a ``more
complicated'' field does not necessarily result in a more complicated
(or ``more chaotic'') motion.

\section{The Null Geodesics}

We will now study the dynamics of the null geodesics in the metric
(\ref{expan}). For null geodesics, $ds^2=0$, and we have:
\beq
E^2g^{tt} + L_z^2g^{\phi\phi} + f(\dot{r}^2 + \dot{z}^2) = 0 ,
\eeq
with $f=-g_{zz}=-g_{rr}$. The effective potential is then given by:
\beq
\frac{E^2}{L_z^2} \equiv \frac{1}{b^2} = V(r,z) = 
	-\frac{g^{\phi\phi}}{g^{tt}} =
	\frac{e^{4\psi}}{r^2} ,
\label{potluz}
\eeq
where $b$ is the impact parameter with respect to the $z$ axis. The
curve $V(r,z) = 1/b^2 = \mbox{const}$ is the boundary of the accessible
regions of the $rz$ plane to a particle having an impact parameter of
$b$. We notice that in the case of massive particles, the effective
potential depends separately on $E$ and $L_z$, while for the case of
massless particles, it 
only depends on the ratio $E/L_z = b$. The equation of motion
for the null geodesics is eq. (\ref{eqmot}) and (\ref{cons}), with $E$
and $L_z$ related by $b=L_z/E$. 

In the case of the dipole field ($Q=O=0$), we find that below a certain value of
the impact parameter $b$ the equipotential curves open, and the orbits
can either fall into the event horizon or escape to infinity. We find,
however, by numerical calculations of pictures of
the basins, which show regular basin boundaries, and by the
computation of the basin boundary dimension, which gives $d=1$ to
within the statistical uncertainty, that the basin boundaries are
regular and the system presents no chaos. This result holds for all values
of $D$ and $b$ we have investigated, and it seems safe to conclude
that massless test particles in the field of a black hole surrounded
by a dipolar material halo move in regular orbits.

When we introduce terms of higher order in the multipolar expansion of
the halo, this situation changes. We illustrate this with a pure
quadrupolar halo. If $Q>0$, there are only two escapes (towards
infinity and towards the event horizon), and the orbits are again
regular. If $Q<0$, however, we have three escape routes (towards the
event horizon, towards $z\tend\infty$ and towards $z\tend-\infty$),
and chaotic behavior arises. This can be seen in fig. 11, where we
show some equipotential curves for $Q=-0.05$, with $D=O=0$. Choosing
$b^2=12.5$, we obtain a picture of the basins by numerical
integration, with the initial conditions in the segment $r=2$,
$|z|<2$. The result is shown in fig. 12a, and an amplification of
several orders of magnitude shows (fig. 12b) that the basin boundary
is fractal. This is further confirmed by the calculation of the basin
boundary box-counting dimension, which yields $d=1.25\pm
0.02$. We have verified that an octopole halo also gives rise to
chaos, as must be the case for multipole terms of higher order. 

Now we make some general remarks on the motion of massless particles
in the metric (\ref{expan}). The absence of stable periodic orbits,
and therefore of a non-zero measure set of confined orbits, was
verified in all cases we have investigated, leading us to believe that
this is a general feature of the motion of zero mass particles in the
black-hole-halo field; we speculate that this may be the case in all
metrics. We also observed that in all cases in which there are only
two escapes (as for instance in the dipolar halo field and in the
quadrupolar field with $Q>0$), the motion of zero rest mass particles
is always regular, as opposed to the motion of material particles in
the same fields. When there are three or more escapes, on the other hand,
chaotic behavior appears. We have found, however, that the motion of
massless particles is always less chaotic than the motion of massive
particles for the same field, for all cases investigated by us.

\section{Conclusions}

We have studied the escape dynamics of a general-relativistic system
consisting of a non-rotating black hole surrounded by a multipolar
halo, which can be thought of as the model o the interior of an
elliptic galaxy with a central massive black hole. For the case of
massive particles, chaos was found in the form of fractal basin
boundaries. We single out the case of a dipolar halo, which has a
Newtonian counterpart that is known to be integrable; the chaos for
this case is thus a result of relativistic corrections to the
dynamics. This is compatible with earlier results obtained with
bounded orbits\ \cite{let1}. Also for this case, we have investigated
the set of trapped orbits with non-zero measure, and we showed that it
is formed by regular orbits, bound to tori in phase space. We have
found that this system is not chaotic for energies above 1, which is
the escape energy for an isolated non-rotating black-hole. The system
appears to be most chaotic (its boundary dimension attains its
highest value) for energies near escape.  

If the halo is composed by a pure quadrupole term, the situation
changes: there can be three escape routes, as opposed to only two
present in the dipole case (this happens if $Q<0$), and in this case
the system is chaotic for arbitrarily large values of the energy, as
far as we could determine. 

In the case of massless particles (null geodesics), we find that the
black hole + dipolar halo system is not chaotic, the basin boundary
between the escapes being regular for all values of the parameters we
have investigated. If we add a quadrupole term to the halo, the motion
becomes chaotic if $Q<0$; the orbits are still regular for
$Q>0$. Terms of higher order also introduce chaos in the
system. Contrary to the case of material particles, we have not found
any stable periodic orbit, with its accompanying non-zero measure set
of confined orbits. We believe the absence of stable periodic orbits
os massless particles is a general feature of axisymmetric static
gravitational fields.

\section*{Acknowledgments}

This research was partially funded by FAPESP and CNPq, and we heartily
thank them.

\newpage
\section*{Figure Captions}

\begin{description}

\item[Figure 1] Level contours of the effective potential for the
dipole field ($Q=O=0$), with $D=3\times 10^{-4}$ and $L_z=3.0$. The
values of $E^2$ for the equipotentials are, from the inside out, 0.93,
0.94, 0.95 and 0.96.

\item[Figure 2] Basin portrait of the section $S$ of the phase space
(see text) for the dipole field, for $D=3\times 10^{-4}$, $L_z=3.0$
and $E^2=0.95$. The black areas correspond to regions of $S$ whose
trajectories fall into the event horizon; white areas correspond to
trajectories that escape to infinity; and white areas correspond to
trajectories that remain trapped inside the confining region. This
figure was calculated on a grid of $400\times 400$ points. (b) is a
magnification of (a).

\item[Figure 3] (a) Poincar\'e section of trapped orbits, with the
surface of section $z=0$,  for $D=3\times 10^{-4}$, $L_z=3.0$ and
$E^2=0.95$. (b) Plot of the fraction of ``uncertain''
points $f(\eps)$ as a function of the separation $\eps$ (see section
3).

\item[Figure 4] Box-counting dimension of the basin boundary as a
function of the dipole strength $D$, for $E=0.95$ and $L_z=3.0$.

\item[Figure 5] Level contours of the effective potential for a
quadrupole field ($D=O=0$) with negative $Q$ ($Q=-4\times 10^{-6}$)
and $L_z=2.6$. The values of $E^2$ for the equipotentials are, from
the inside out, 0.93, 0.94, 0.95 and 0.96.

\item[Figure 6] Basin portrait for the quadrupole field, with
$Q=-4\times 10^{-6}$ and $L_z=2.6$. Black areas
denote regions whose trajectories fall into the event horizon; gray
areas correspond to trajectories that escape towards $z\tend +\infty$;
and white areas correspond to trajectories that escape towards $z\tend
-\infty$. (b) is a magnification of (a).

\item[Figure 7] Time of escape versus the velocity angle ($Q=-4\times
10^{-6}$, $L_z=2.6$), for $r=25$
and $z=0$. (b) is a magnification of (a).

\item[Figure 8] (a) $I_1$, (b) $I_2$, (c) $I_3$, (d) $R_2$, for $Q=-4\times
10^{-6}$, $L_z=2.6$ and $D=O=0$.

\item[Figure 9] Level contours of the effective potential for the
classical quadrupole field with $Q=-4\times 10^{-6}$ and
$L_z=2.6$. The values of $E^2$ for the equipotentials are, from the
inside out, 0, -0.01, -0.02 and -0.03.

\item[Figure 10] Level contours of the effective potential for the
field with $D=0$, $Q=-4\times 10^{-6}$, $O=-10^{-7}$ and $L_z=2.6$.
The values of $E^2$ for the equipotentials are, from the inside out,
0.94, 0.95, 0.96 and 0.97.

\item[Figure 11] Level contours of the effective potential for null
geodesics for the quadrupole field  $Q=-0.05$, $D=O=0$. The values
of $b^2$ ($b$ is the impact parameter with respect to the
symmetry axis) are, from the inside out, 10.0, 15.0 and 20.0.

\item[Figure 12] Basin portrait for null geodesics with $Q=-0.05$,
$D=O=0$ and $b^2=12.5$. Areas in black correspond to trajectories that
fall into the event horizon; areas in gray correspond to trajectories
that escape towards $z\tend +\infty$, and areas in white denote
trajectories that escape to $z\tend -\infty$.

\end{description}

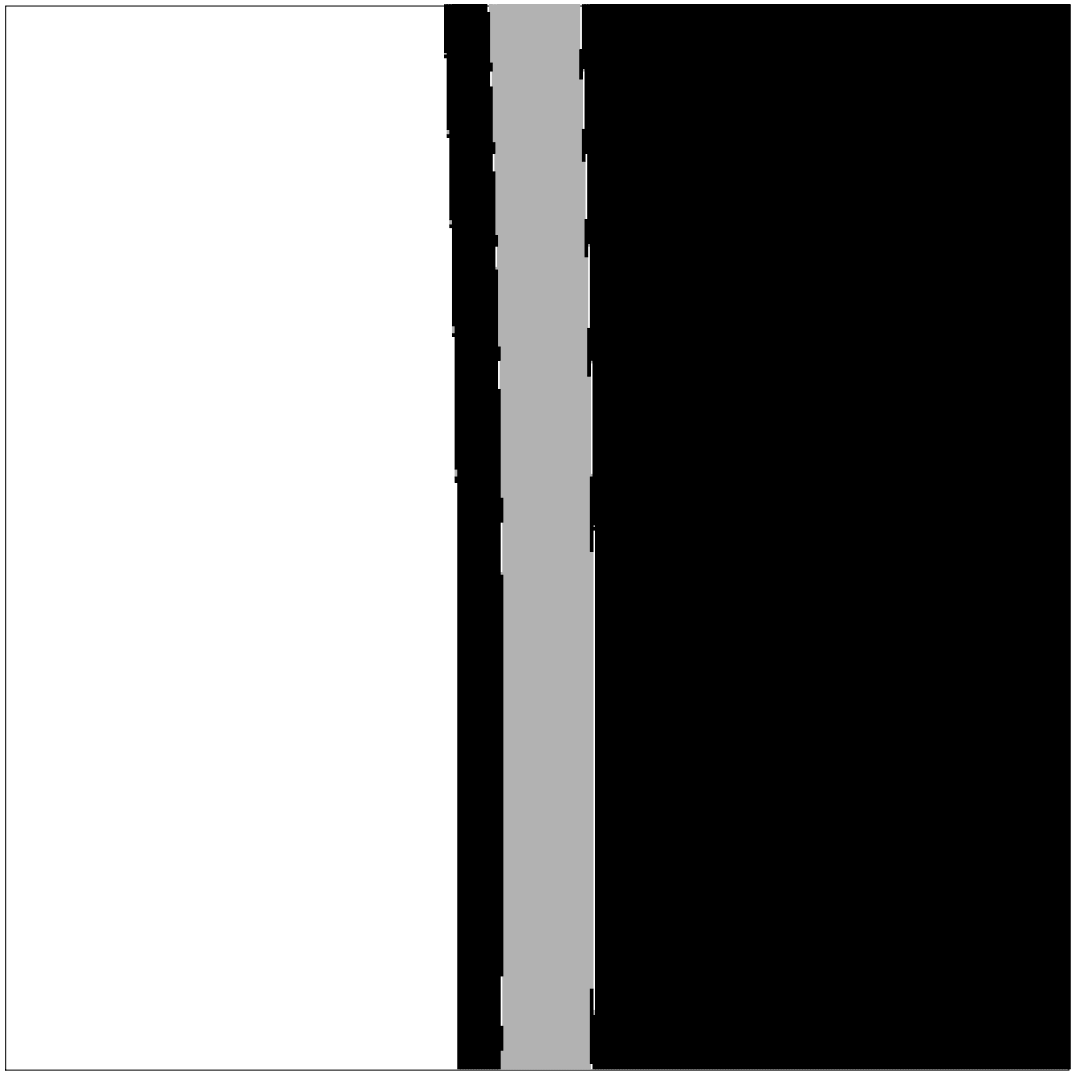
\end{document}